\documentclass[aps,superscriptaddress,twocolumn,nopreprintnumbers,floatfix]{revtex4-1}

\usepackage{amssymb}
\usepackage{amsmath}
\usepackage{graphicx}
\usepackage{dcolumn}
\usepackage{color,units}
\usepackage{lineno}
\usepackage{xspace}
\usepackage{mathtools}
\usepackage{physics}
\usepackage{tensor}
\usepackage{bm}
\usepackage{lipsum}
\usepackage[normalem]{ulem}
\usepackage[plainpages=false, colorlinks=true, anchorcolor=blue, linkcolor=blue, citecolor=blue, bookmarks=false]{hyperref}

\newcommand{\bilby}{{\sc Bilby}\xspace}

\newcommand{\Z}{\mathcal{Z}}

\newcommand{\V}{\mathcal{V}}
\renewcommand{\L}{\mathcal{L}}
\newcommand{\lmax}{\ell_\textrm{max}}

\newcommand{\SPA}{School of Physics and Astronomy, Monash University, Clayton VIC 3800, Australia}
\newcommand{\OzGravMonash}{OzGrav: The ARC Centre of Excellence for Gravitational Wave Discovery, Clayton VIC 3800, Australia}

\begin{document}

\title{Searching for anisotropy in the distribution of binary black hole mergers}

\author{Ethan Payne}
\email{ethan.payne@ligo.org}
\affiliation{\SPA}
\affiliation{\OzGravMonash}

\author{Sharan Banagiri}
\affiliation{School of Physics and Astronomy, University of Minnesota, Minneapolis, MN 55455, USA}

\author{Paul D. Lasky}
\author{Eric Thrane}
\affiliation{\SPA}
\affiliation{\OzGravMonash}

\date{\today}

\begin{abstract}
The standard model of cosmology is underpinned by the assumption of the statistical isotropy of the Universe.
Observations of the cosmic microwave background, galaxy distributions, and supernovae, among other media, support the assumption of isotropy at scales $\gtrsim 100$\,Mpc. 
The recent detections of gravitational waves from merging stellar-mass binary black holes provide a new probe of anisotropy; complementary and independent of all other probes of the matter distribution in the Universe.
We present an analysis using a spherical harmonic model to determine the level of anisotropy in the first LIGO/Virgo transient catalog.
We find that the ten binary black hole mergers within the first transient catalog are consistent with an isotropic distribution.
We carry out a study of simulated events to assess the prospects for future probes of anisotropy.
Within a single year of operation, third-generation gravitational-wave observatories will probe anisotropies with an angular scale of $\sim36^\circ$ at the level of $\lesssim0.1\%$.

\end{abstract}

\maketitle

\section{Introduction}

Two key assumptions in the standard Lambda cold dark matter ($\Lambda$CDM) model of modern cosmology are the isotropy and homogeneity of the Universe \cite{efstathiou1990lcdm, peebles1982large, peebles2003review}. 
Since $\Lambda$CDM cosmology underlies our current understanding of the Universe, rigorous tests are required to validate these fundamental assumptions.
The Universe has been observed to be isotropic at scales $\gtrsim 100$\,Mpc, with smaller-scale deviations considered to be \textit{statistically isotropic} on an overall homogeneous structure 
produced by phenomena such as baryon acoustic oscillations or gravitational interactions \cite{hogg2005cosmic, yadav2005testing, marinoni2012scale}. 
Verifying the lack of large scale anisotropy in the Universe is important for the validity of the  $\Lambda$CDM cosmology.

Evidence for large-scale isotropy of the Universe has been presented through numerous observations of various sources \cite{smoot1992structure, fixsen1996cosmic, wu1999large, blake2002velocity, marinoni2012scale, meegan1992spatial, scharf1999evidence, akrami2019planck, planck2014XXVII}. 
The most stringent measurements on the anisotropy of the Universe come from the cosmic microwave background, which generally show small-scale statistical deviations on the order of $10^{-4}$---$10^{-5}$~\cite{planck2014XXVII, akrami2019planck}.
Meanwhile, multiple studies hint at the existence of deviations from isotropy at large angular scales in the cosmic microwave background~\cite{eriksen2007hemispherical, vielva2004detection}, supernovae~\cite{schwarz2007isotropy, colin2019evidence} and galaxy~\cite{javanmardi2017anisotropy, migkas2020probing} distributions, as well as large bulk flows~\cite{watkins2009consistently, kashlinsky2010new}. 
While these inferences are speculative~\cite[e.g.,][]{aguirre1998dust,itoh2010dipole, Rubin2020}, they can be supported or contradicted by independent measures of the Universe's isotropy using gravitational-wave observations. 

Studies of the gravitational-wave stochastic background have placed limits of anisotropy from unresolved sources \cite{thrane2009,directional_limits2017, directional_limits2019, Renzini2019improved, Ain2018folded, banagiri2020measuring, mukherjee2020time}, however
observations of gravitational waves from resolved binary black hole (BBH) mergers present another tool to probe anisotropy.
Prior to the third observing run, Advanced LIGO (aLIGO)~\cite{aligo} and Advanced Virgo (aVirgo)~\cite{Virgo} released the details for ten binary black holes over the first (O1) and second (O2) observing runs \cite{abbott2019gwtc}.
These observations are collated in the first gravitational-wave transient catalog, GWTC-1. 
With the third observing run (O3) complete, the total number of gravitational-wave BBH merger candidates has increased to more than 50~\cite{gracedb}.
Furthermore, the addition of aVirgo for the entirety of O3 has already resulted in many well-localized sources (e.g. \cite{abbott2017gw170814}), providing further motivation for utilizing BBH mergers to study the anisotropy of the Universe. 

In this paper, we use LIGO/Virgo data to probe anisotropies in the distribution of BBH mergers, taking care to handle selection bias associated with the detection of gravitational-wave sources.
We explore the future prospects of anisotropy measurements with gravitational waves.
In Ref. \cite{stiskalek2020isotropy}, a HEALPix~\cite{gorski2005healpix} basis was adopted to parameterize the anisotropy of binary black hole mergers.
In Ref.~\cite{cavaglia2020two}, a two-dimensional correlation function was implemented, where the angular power spectrum was inferred assuming statistical isotropy of sources (see also \cite{vijaykumar2020probing,mukherjee2020,mukherjee2018beyond}).
Both analyses find results consistent with isotropy. 
In contrast, our method utilizes a spherical harmonic basis to define a probability distribution, providing results in terms of typical spherical harmonic functions.
Our results are qualitatively similar to those of~\cite{stiskalek2020isotropy, cavaglia2020two}.
However, there are some potential advantages to the spherical harmonic approach: the method lends itself straightforwardly to comparison with results from the cosmic microwave background, and it could be argued that at least some plausible deviations from anisotropy are more clearly visible in the spherical harmonic basis than the pixel basis, which emphasizes hot spots.

The remainder of the manuscript is structured as follows. In Sec. \ref{sec:method}, we outline a method of hierarchically analysing binary black hole mergers to determine the level of anisotropy, including a parameterization of the distribution of BBH merger events in spherical harmonics and a discussion of the observational selection biases. The analysis of the previously observed BBH events in GWTC-1 is presented in Sec. \ref{sec:gwtc1}. In Sec. \ref{sec:demonstration}, we look to the near future by demonstrating the recovery of an isotropic and anisotropic universe with simulated gravitational-wave events.
We conclude with an investigation into the future of using gravitational waves as an indicator for anisotropy in the matter distribution of the Universe and the associated implications.

\section{Methodology} \label{sec:method}
In order to test the assumption of an isotropic universe with gravitational-wave observations, we employ a spherical harmonic basis. 
We do not employ a \textit{linear} spherical harmonic basis expansion as this can result in a negative probability density. Instead, we employ a \textit{non-linear} spherical harmonic basis to guarantee positivity (see also~\cite{taylor2020bright}).
We write the probability density function for mergers $\pi(\bm{\Omega})$ as
\begin{equation}
    \pi(\bm{\Omega}|\{b_{\ell m}\}) = \frac{1}{\sum_{\ell m} \abs{b_{\ell m}}^2}\Big(\sum_{\ell m} b_{\ell m}Y_{\ell m}(\bm{\Omega})\Big)^2, \label{eq:prior}
\end{equation}
where
\begin{align}
\sum_{\ell m} \equiv \sum_{\ell=0}^{\ell_\textrm{max}}\sum_{m=-\ell}^\ell.
\end{align}
Here, $\bm{\Omega}$ is the sky position unit vector, $Y_{\ell m}(\bm{\Omega})$ are spherical harmonics, and $\{b_{\ell m}\}$ is the set of coefficients, up to $\ell =\lmax$, that parameterize the anisotropy of the distribution.
The total number of parameters for a given model is $\lmax(\lmax+2)$, where we set the monopole term to $b_{00}=1$ without loss of generality. The maximum resolvable $\ell$ is related to the angular size of individual events and the angular separation within the population. Analyses limited by the angular resolution of individual detections cannot resolve spherical harmonics with $\ell \gtrsim 30$~\cite{cavaglia2020two}. However, we also need to consider the effect of the number of events limiting our analysis. If the minimum angular separation between detected events exceeds the typical angular scale of the $(\ell,m)$ spherical harmonic, $180^\circ/\ell$, then the spherical harmonic can never be resolved. For ${\cal O}(100)$ events, the $\ell=5$ harmonics can be resolved. 
Furthermore, by restricting $\lmax$, we also target the larger angular sizes of speculative anisotropies~\cite{eriksen2007hemispherical, vielva2004detection, schwarz2007isotropy, colin2019evidence, javanmardi2017anisotropy, migkas2020probing, watkins2009consistently, kashlinsky2010new}.
Therefore, in this manuscript, we restrict our analysis to $\lmax \leq 5$.

Note that $\pi(\bm\Omega)$ is given not as a linear combination of spherical harmonic coefficients, but as the square of this sum.
Coupled with, 
\begin{align}
    b_{\ell, -m} \equiv (-1)^mb_{\ell m}^* ,
\end{align}
implying the $b_{\ell m}$ are complex for $m\neq0$ and that the $b_{\ell 0}$ are real,
we ensure that the probability density function is positive definite.
Since the $m<0$ modes are uniquely determined from the $m>0$ modes, we consider the set of parameters (which we sample over to be) $\{ b_{\ell m}\}$ with $\ell\geq 1,m \geq 0$, but all summations occur over the full range of allowed $m$'s.

The orthogonality of spherical harmonics allows us to express $\pi(\bm{\Omega})$ as a linear combination of spherical harmonics
\begin{equation}
    \pi(\bm{\Omega}) = \sum_{LM}a_{LM}Y_{LM}(\bm{\Omega}). \label{eq:typical_sph}
\end{equation}
The $a_{\ell m}$'s are related to the $b_{\ell m}$ by Clebsch-Gordan coefficients:
\begin{equation}
    a_{LM} = \frac{1}{\sum_{\ell m} \abs{b_{\ell m}}^2}\sum_{\ell m}\sum_{\ell' m'} \beta_{\ell m, \ell ' m'}^{LM}b_{\ell m}b_{\ell' m'},\label{eq:alms}
\end{equation}
where
\begin{equation}
    \beta_{\ell m, \ell ' m'}^{LM} \equiv 
        \displaystyle\sqrt{\frac{(2\ell+1)(2\ell'+1)}{4\pi(2L+1)}}C^{LM}_{\ell m, \ell'm'}C^{L0}_{\ell0, \ell'0}.
\end{equation}
Here, $C^{LM}_{\ell m, \ell'm'}$ is the set of well-known Clebsch Gordan coefficients \cite{wigner1931gruppentheorie, majumdar1958}. 
The reconstructed values of $a_{\ell m}$'s extend up to $2\lmax$. Throughout the manuscript, $\lmax$ refers to the $\ell$'s of the $b_{\ell m}$'s.

The relation of the $b_{\ell m}$ to the $a_{\ell m}$ can be further simplified in the limit of small deviations from isotropy, $b_{\ell m}/b_{00} \ll 1$, to
\begin{align}
    a_{LM} &= \frac{b_{LM}}{\sqrt{\pi}b_{00}} + \mathcal{O}(b_{\ell m}^2/b_{00}^2).\label{eq:simplications}
\end{align}
where the monopole is always $a_{00} = 1/\sqrt{4\pi}$.
Equation \eqref{eq:simplications} is not used for any calculations in this manuscript, but it is useful for understanding the relationship between the $b_{\ell m}$ and $a_{\ell m}$.
With this model, we aim to measure the posterior distributions of the $a_{\ell m}$'s, where non-zero values imply the existence of an anisotropic BBH merger distribution.

In order to fit the $b_{\ell m}$'s, we employ  hierarchical inference~\cite{gelman2013bayesian, thrane2019introduction}.
We marginalize over $\bm\Omega$ to obtain ``hyper-likelihoods'' for the data given the $b_{\ell m}$.
The hyper-likelihood for a set of data from $N$ events, $\{d_i\}$, is
\begin{widetext}
\begin{equation}
    \L(\{d_i\} | \{b_{\ell m}\}) = \frac{1}{p_\textrm{det}(\{b_{\ell m}\} | N)}\prod_i^N \frac{\Z_\textrm{iso}(d_i)}{n_i}\sum_k^{n_i}\frac{\pi(\bm{\Omega}^k_i | \{b_{\ell m}\})}{\pi(\bm{\Omega}^k_i |\textrm{iso})}. \label{eq:hyperlikelihood}
\end{equation}
\end{widetext}
Here, $\{b_{\ell m}\}$ are the spherical harmonic coefficients for the anisotropic model in Eq.~\eqref{eq:prior}, up to $\ell=\ell_\textrm{max}$, and $p_\textrm{det}(\{b_{\ell m}\}| N)$ is the probability of detection given a set of hyper-parameters. 
The isotropic probability distribution is defined as $\pi(\bm{\Omega} |\textrm{iso})$, defined as $b_{\ell m} = 0$ for $\ell \geq 1$,
${\cal Z}_\text{iso}(d_i)$ is the $i$th event's evidence assuming the isotropic sky location probability distribution, and $n_i$ is the total number of samples in the $i$th event's posterior distribution.
We note that this implementation does not account for false positive detections of BBH mergers. 
Procedures to include the possibility of falsely identified sources rely heavily on the probability that a given event was astrophysical~\cite{galaudage2019}.
These probabilities vary between pipelines~\cite{abbott2019gwtc}, and therefore we have assumed all detections are astrophysical in origin for simplicity.
The detection probability term depends on the selection bias of the interferometers and is discussed in depth below.

We use the samples and Bayesian evidence from astrophysical inference of gravitational-wave events.
By default, the evidence for event $i$ is calculated assuming an isotropic distribution, ${\cal Z}_\text{iso}(d_i)$.
The total evidence for isotropy, $\Z_\textrm{iso}$, is simply the product of all individual event evidences. 
The hyper-posterior is 
\begin{equation}
    p(\{b_{\ell m}\}| \{d_i\}) = \frac{\L(\{d_i\} | \{b_{\ell m}\}) \pi(\{b_{\ell m}\})}{\Z_\textrm{ani}},
\end{equation}
where $\Z_\textrm{ani}$ is the evidence for the anisotropic model,
\begin{equation}
    \Z_\textrm{ani} = \int \L(\{d_i\} | \{b_{\ell m}\}) \pi(\{b_{\ell m}\})\,\dd\{b_{\ell m}\},
\end{equation}
and $\pi(\{b_{\ell m}\})$ is our hyper-prior.

To present the predicted BBH merger sky position probability density function, we calculate the posterior predictive distribution (PPD)
\begin{align}
    p_{\text{PPD}}(\bm\Omega) = & \int \pi(\bm\Omega |\{b_{\ell m}\}) p(\{b_{\ell m}\} |\{d_i\})\dd\{b_{\ell m}\}.
\end{align}
The PPD predicts the angular distribution given the previously observed events and can be approximated numerically from the hyper-posterior samples.
We note that although the PPD allows visualisation of the average posterior, it does not capture the full details of the posterior, hiding the full range of possible distributions.
We use the Bayes factor 
\begin{equation}
    \textrm{BF}^\textrm{ani}_{\textrm{iso}} \equiv \frac{\Z_\textrm{ani}}{\Z_{\textrm{iso}}},
\end{equation}
to determine if one model is preferred relative to another.
We adopt a threshold of $\ln \textrm{BF}^\textrm{ani}_{\textrm{iso}}=8$ for a statistically significant signature of anisotropy. 

Gravitational-wave observatories do not observe the sky with an isotropic sensitivity. 
The detector response is characterized by an ``antenna factor,'' which varies depending on the sky position of the source relative to the L-shaped geometry of the interferometer~\cite{schutz2011networks}. 
During a sidereal day, this antenna pattern function is swept about right ascension. 
Additionally, gravitational-wave observatories are more likely to be operating during the night due to reduced anthropogenic activity \cite{chen2017observational}. 
Furthermore, even when the observatories are operating during the day, the noise power spectral density is on average higher than at night. 

In order to account for selection effects, we first calculate the detection probability as a function of the sky position of the source, denoted $p_\textrm{det}(\bm{\Omega})$.
We calculate the detection probability by determining the fraction of binaries detected at a particular distance, time, and sky position \cite{tiwari2018constraining},
\begin{equation}
    p_\textrm{det}(\bm\Omega) = \frac{1}{\V_{\textrm{tot}}T}\int_{t_0}^{t_0+T} \int_0^{z_\textrm{max}} \dv{V_c}{z}\frac{1}{1+z}f(z, \bm\Omega, t)\,\dd z\dd t\label{eq:pdet_radec} .
\end{equation}
Here, $\V_\textrm{tot}$ is the total volume of our population model
\begin{equation}
    \V_{\textrm{tot}} = \int_0^{z_\textrm{max}} \dv{V_c}{z}\frac{1}{1+z}\,\dd z,
\end{equation}
The variable, $V_c$ is the comoving volume, $f(z, \bm\Omega, t)$ is the selection function determining the fraction of binaries observed at a given sky position, time, and redshift $z$. 
Finally the variables $t_0$ and $T$ correspond to the start of the observing period and the duration respectively.
We marginalize over time within the network to ensure that the detection probability accounts for all possible times during observing runs when a BBH merger event could be detected.

The selection function is calculated by simulating gravitational-wave events from an astrophysically motivated population distribution, $\pi_{\textrm{pop}}(\bm\theta)$, and determining the fraction of detected events from their SNR.
For each simulated event, we calculate the signal-to-noise ratio (SNR) for individual observatories $\rho_\text{ifo}$ and for the entire network $\rho_\text{net}$. 
The expectation value of the matched-filter SNR within a single interferometer for a particular event is \cite{cutler1994gravitational},
\begin{equation}
    \rho^2_{\textrm{ifo}} \equiv 4\int_0^{\infty}\frac{|F_+(\bm\Omega, \iota, \psi)h_+(f)+F_\times(\bm\Omega, \iota, \psi) h_\times(f)|^2}{S_n(f)}\,\dd f.
\end{equation}
This is often referred to as the ``optimal'' signal-to-noise ratio. 
Here, $F_{+,\times}$ denote the antenna pattern function for plus and cross polarizations, $h_{+,\times}$ are the associated gravitational-wave strains for each polarization, and $S_n(f)$ denotes the noise power spectral density (PSD) of the interferometer.
For a network with $N_\textrm{ifo}$ interferometers, the network signal-to-noise ratio is
\begin{equation}
    \rho^2_\textrm{net} = \sum_{i}^{N_\textrm{ifo}}\rho^2_{i}.
\end{equation}
We adopt the criteria that a signal is detected if $\rho_{\textrm{ifo}}>8$ or $\rho_{\textrm{net}}>12$. 

For the population distribution, we assume a power law distribution for the source frame masses of the binary systems as given in Ref. \cite{abbott2019binary} with a primary mass power law of $\pi_{\textrm{pop}}(m_1) \propto m_1^{-1.6}$, and a minimum and maximum mass of $7.9$\,$M_\odot$ and $42.0$\,$M_\odot$, respectively. 
The mass ratio distribution is assumed to be $\pi_{\textrm{pop}}(q) \propto q^{6.7}$. 
We assume a uniform prior in dimensionless spin magnitude between (0,0.9) with isotropic spin orientations.
We employ standard priors for extrinsic parameters with a uniform-in-source-frame volume prior for distance.
The choice of the population distribution predominantly affects the overall scaling of the detection probability. In contrast, the configuration of the detector network establishes its angular structure.

We use Monte Carlo integration to calculate the integral in Eq.~\eqref{eq:pdet_radec}, which yields $p_\text{det}(\bm\Omega)$.
We generate $10^4$ realizations of the noise power spectral density for observatories in our network.
Each realization is assigned a random time from the O1/O2 observing runs.
Using this set of PSDs, we calculate the detection probability,  $p_\textrm{det}(\bm\Omega, t_i)$, as a function of sky position at each time $t_i$. 
The calculation is carried out using a 3072-pixel HEALPix \cite{gorski2005healpix} grid, which provides a sufficiently fine resolution relative to the angular scale of $\ell\leq5$ spherical harmonic functions.
The fraction of events from the BBH injections that exceed the signal-to-noise ratio detection threshold at each sky position determines $p_\textrm{det}(\bm\Omega, t_i)$.
All realizations of the detection probability at different times are then averaged to determine $p_\textrm{det}(\bm\Omega)$.

\begin{figure}
    \centering
    \includegraphics[width=\linewidth]{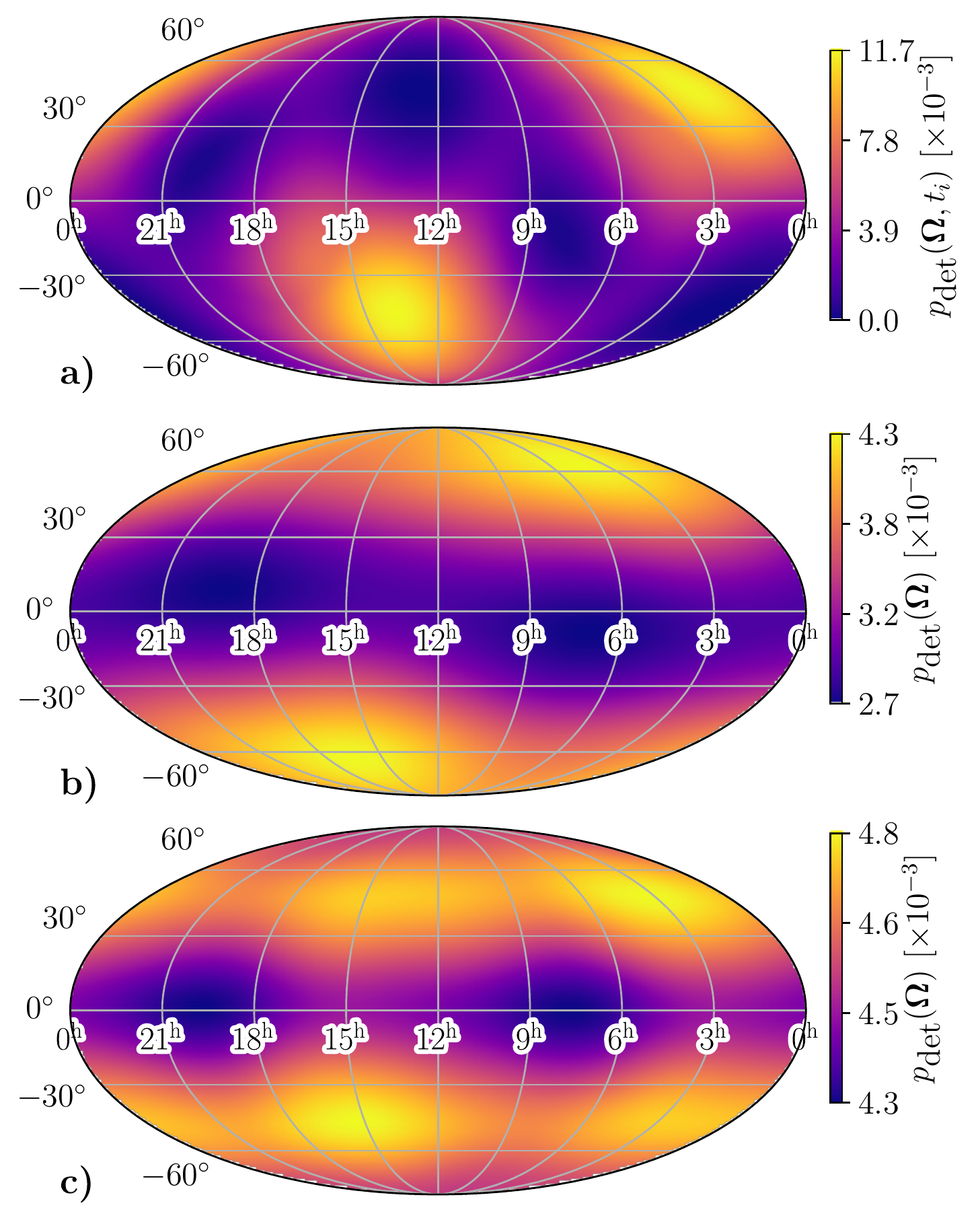}
    \caption{Probability of detection for a number of different scenarios. (a) Detection probability as a function of sky position at 10:10:20 14\textsuperscript{th} September, 2015. 
    (b) Detection probability over the entire first observing run.
    (c) Detection probability over the first and second observing runs.}
    \label{fig:pdet_ra_dec}
\end{figure}

Figure \ref{fig:pdet_ra_dec} presents the detection probabilities (a) for a particular instant in time, (b) marginalized over the first observing run, and (c) marginalized over both the first and second observing runs. Then detection probability at any given time, as seen in Fig. \ref{fig:pdet_ra_dec}(a), is dominated by the antenna pattern functions of the interferometers.
The marginalized detection probability skymaps in Fig. \ref{fig:pdet_ra_dec}(b) and \ref{fig:pdet_ra_dec}(c), present more non-trivial behaviour due to the interplay of the antenna factors with the time-dependence of the detectors' performance. 
In general, the probability of detection is smeared over the sky over extended periods of times, suppressing the lobe structure due to the antenna pattern function. 
Within our analysis of GWTC-1, only the combined detection probability is utilized.
However the O1 detection probability is presented to highlight the strong selection bias present in O1 analysis. 
This was primarily due to both LIGO interferometers preferring a mid-declination due to their antenna pattern function maxima, and the consistency of a diurnal cycle during the first observing run \cite{chen2017observational}. 
These features are less clear in the combined detection probability due to the duration of the second observing run and improved duty cycle. 
To determine the detection probability as a function of the hyper-parameters, we compute
\begin{equation}
    p_\textrm{det}(\{b_{\ell m}\} | N) \propto \Big(\int p_\textrm{det}(\bm\Omega)\pi(\bm\Omega | \{b_{\ell m}\}) \,\dd\bm\Omega\Big)^N,
\end{equation}
where $N$ is the number of events observed \footnote{This implicitly assumes a uniform-in-log prior on the merger rate \cite{thrane2019introduction}. This is consistent with the population inference of GWTC-1 undertaken in Ref. \cite{abbott2019binary}.}.

We implement the method outlined above with the Bayesian inference library, \bilby \cite{ashton2019bilby}, using the nested sampling algorithm, {\tt \sc Dynesty} \cite{speagle2019dynesty}. 
To implement the spherical harmonic model in the nested sampler, we sample in $b_{\ell m}/b_{00}$ for $\ell > 0$ without loss of generality. 
To first order, $b_{\ell m}/b_{00} \approx a_{\ell m}/2a_{00}$, implying sampling in $b_{\ell m}/b_{00}$ directly presents the degree of anisotropy in the distribution. 
For each spherical harmonic coefficient, we define a uniform hyper-prior on the magnitude from 0 to $1/\sqrt{2}$.
For $m=0$ modes, this applies to the coefficient $b_{\ell m}/b_{00}$, sampling uniformly over $[-1/\sqrt{2},1/\sqrt{2}]$. 
For $m>0$ modes, we sample in the magnitude and phase of $b_{\ell m}/b_{00}$ using the uniform magnitude prior and a uniform phase prior.
The prior distributions are constructed to fully span the $a_{\ell m}$ parameter range for $\ell \le \lmax$, while limiting the magnitude of $a_{\ell m}$'s for $\ell > \lmax$. 
This significantly reduces multi-modality of the likelihood, ensuring improved convergence of the nested sampling algorithm.

\section{GWTC-1} \label{sec:gwtc1}
We apply the above methodology using $\lmax=1$ through $5$ to data from GWTC-1~\cite{abbott2019gwtc}, 
utilizing the samples from~\cite{bilby_gwtc1} and reweighting to use the population distribution as our prior. 
We down-sample to 3000 individual samples for each event to make the calculation more tractable. 
We apply the detection probability skymap calculated for the combination of O1 and O2; see Fig. \ref{fig:pdet_ra_dec}(c).
The Bayes factors between anisotropy and isotropy ($b_{\ell m} = 0$ for all $\ell >0$) for the different models are reported in Table~\ref{tab:gwtc1_logbf}.

\bgroup
\def\arraystretch{1.5}
\begin{table}[t!]
    \centering
    \caption{Natural log Bayes factors for each of the different anisotropic models used to analyse GWTC-1 comparing the anisotropic hypothesis to the isotropic hypothesis. We find no evidence for a preference of an anisotropic model over isotropy.}
    \begin{tabular}{c|c}\hline\hline
        $\ell_\textrm{max}$ & $\ln\textrm{BF}^{\textrm{ani}}_{\textrm{iso}}$ \\ \hline
        1 & $-0.45$  \\
        2 & $-0.32$ \\
        3 & $-0.14$ \\
        4 & $-0.15$ \\
        5 & $-0.12$ \\ \hline\hline
    \end{tabular}
    \label{tab:gwtc1_logbf}
\end{table}
\egroup

\begin{figure*}[t!]
    \centering
    \includegraphics[width=0.9\linewidth]{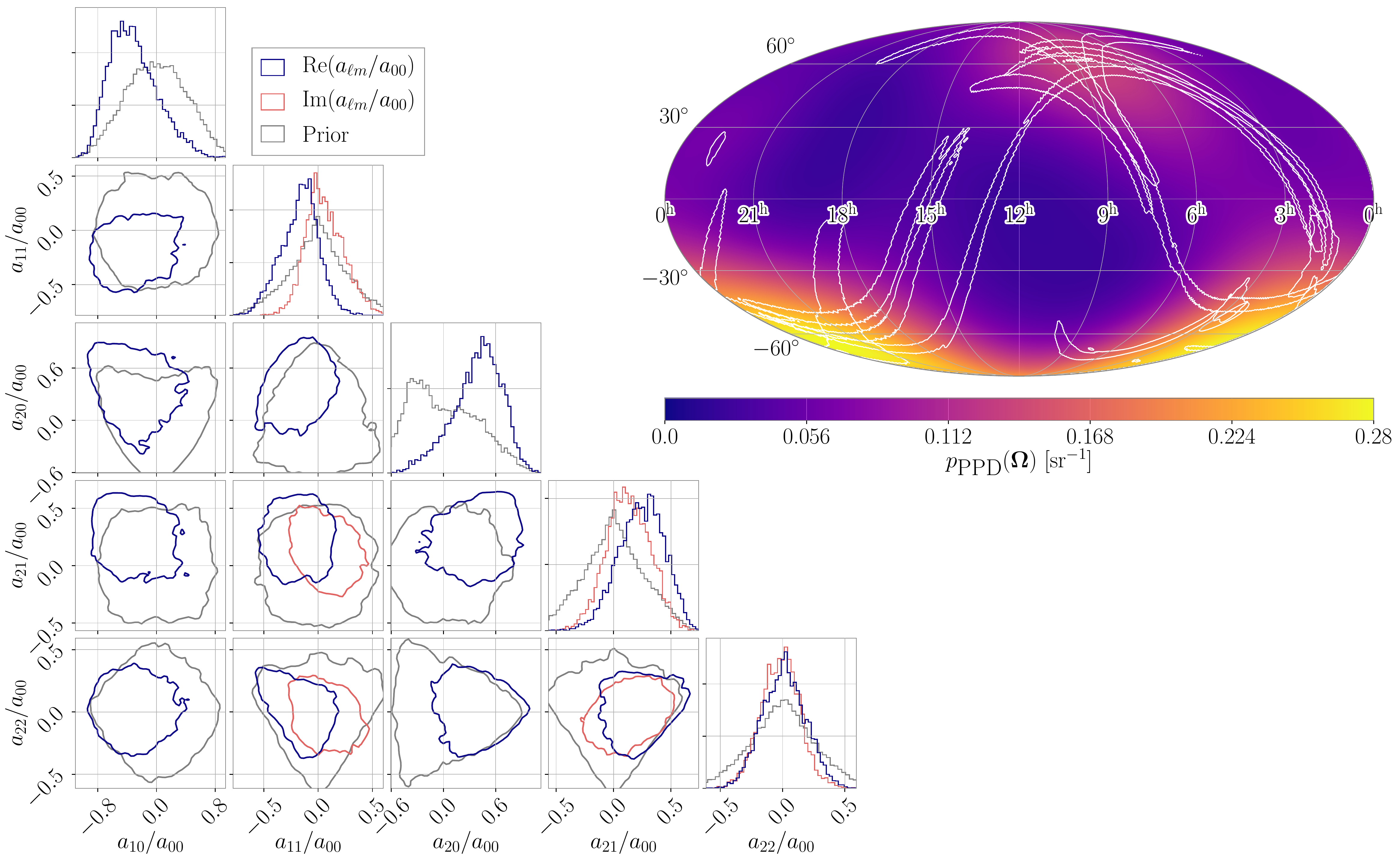}
    \caption{Posterior probability distribution showing the $95\%$ credible intervals (left) and posterior predictive distribution (right) of the analysis of GWTC-1 with the $\lmax=2$ anisotropy model. The posterior distribution includes the real and imaginary components for the $m\neq0$ spherical harmonic coefficients. 
    The prior distribution (grey) is determined from the reconstruction of $a_{\ell m}$s from the uniform $b_{\ell m}$ priors. 
    This results in notable features such as the $a_{20}$ posterior peaking away from $a_{20}=0$. 
    These effects are suppressed with increasing $\lmax$.
    While the corner plot does not meaningfully exclude zero, the posterior predictive distribution does present regions of higher probability as Ref. \cite{stiskalek2020isotropy} has also observed. However, isotropy is not excluded by this result and the Bayes factor for $\lmax=2$ anisotropy does not prefer either model. The white contours in the posterior predictive distribution correspond to the $95\%$ confidence intervals for the locations of all BBH merger events in GWTC-1.}
    \label{fig:corner_ppd_gwtc1}
\end{figure*}

We find the isotropic model is preferred over the anisotropic models with the Bayes factors ranging from $\ln\textrm{BF}=-0.45$ for $\ell_\textrm{max}=1$ to $-0.12$ for $\ell_\textrm{max}=5$. 
We present the posterior distributions of $a_{\ell m}/a_{00}$ for $\ell \leq 2$ for the $\lmax=2$ anisotropy model in Fig. \ref{fig:corner_ppd_gwtc1}. The real and imaginary components are shown for the $m\neq0$ spherical harmonic coefficients.
The posterior distributions for the other models are presented in Appendix~\ref{app:a}.
The posterior distribution for all anisotropy models differ from the prior distribution, indicating that the ten events in GWTC-1 do provide some information about the overall sky distribution of events. 
However, the posterior distributions are still consistent with an isotropic universe ($a_{\ell m} = 0$ for $\ell \geq 1$). 
There this support for all $a_{\ell m} = 0$ present to at least the $95\%$ level.

We also present the posterior predictive distribution for the $\lmax=2$ anisotropic analysis in Fig. \ref{fig:corner_ppd_gwtc1}.
Although there is no preference for the $\lmax=2$ anisotropic model, the posterior predictive distribution illustrates what the anisotropy may look like with the anisotropic model if it was detected in GWTC-1.
These results are qualitatively similar to Ref. \cite{stiskalek2020isotropy}, where similar preferences for BBH sky positions are observed using a HEALPix \cite{gorski2005healpix} based model with 12 individual pixels, and three rotation angles. They report mild support for isotropy. 

\begin{figure*}
    \centering
\includegraphics[width=\linewidth]{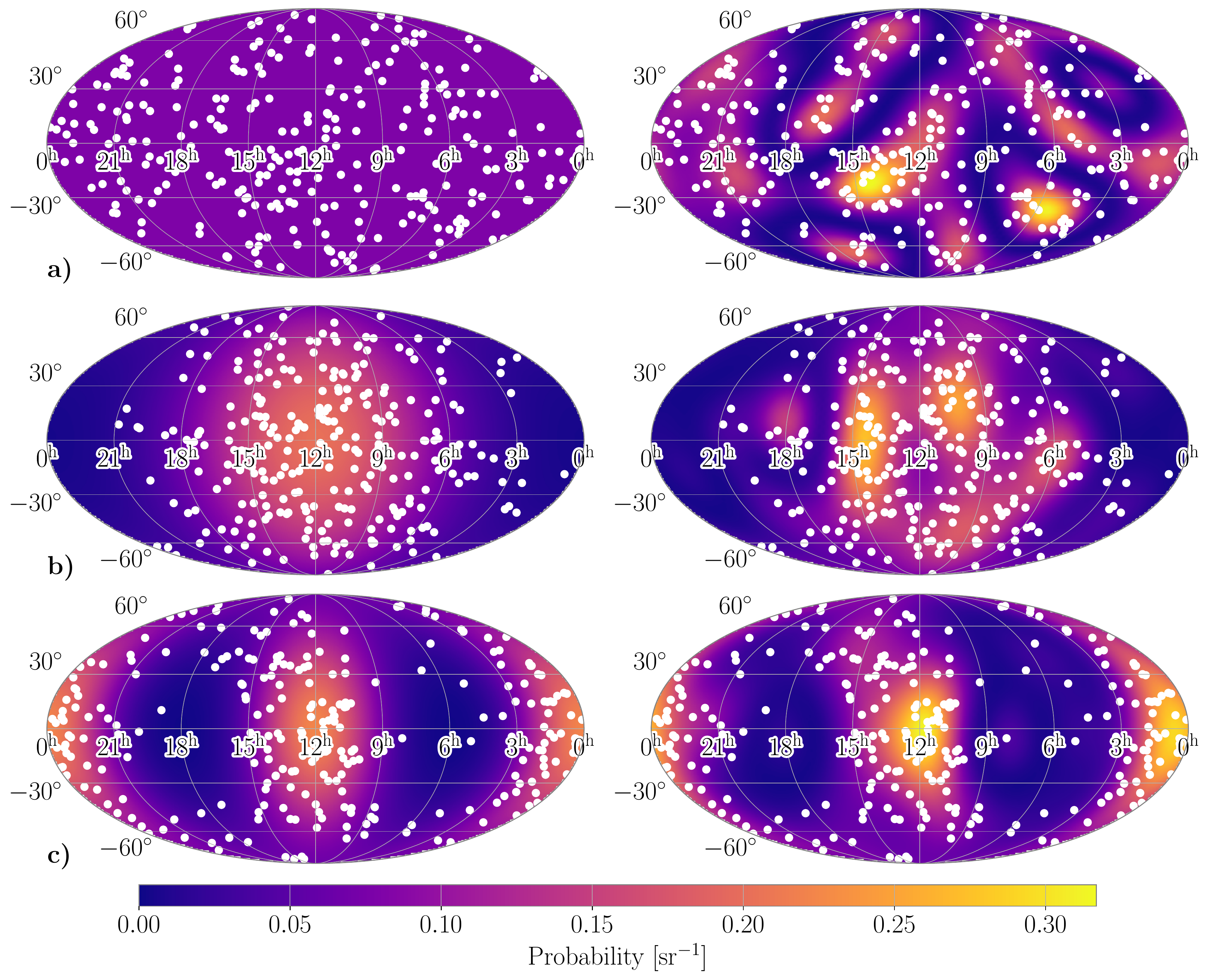}
    \caption{Injected binary black hole merger sky distribution (left) and the associated recovered posterior predictive distribution (right), for (a) an isotropic universe, (b) a universe with an $a_{11}/a_{00} = 0.51$ dipolar anisotropy, and (c) a universe with an $a_{22}/a_{00} = 0.51$ quadrupolar anisotropy. The sky distribution is recovered from 269 events drawn from the population prior. All events are drawn such that they exceed the detection threshold of the three detector network. We determine the detection probability as a function of signal position to remove this selection bias. We recover an approximately similar sky distribution to the injected distribution, demonstrating the accurate recovery of anisotropy.}
    \label{fig:injection_recovery}
\end{figure*}

\section{Simulated event analysis}\label{sec:demonstration}
In this section, we assess how future observations will be able to probe anisotropy of compact binary mergers.
We consider a network consisting of two aLIGO observatories and one aVirgo observatory, all operating at design sensitivity.
We assume this network operates with a 100\% duty cycle. 
The detection probability skymap is calculated following an identical procedure to the analysis of GWTC-1, except there is no downtime and we assume stationary Gaussian noise. 
We simulate gravitational-wave signals from binary black hole mergers with parameters distributed according to the aforementioned model over the course of one year of three-detector coincident operation.

We create simulated data for three angular distributions: isotropic, dipolar ($a_{11}/a_{00} =0.51$), and quadrupolar ($a_{22}/a_{00} =0.51$) distributions with all other $a_{\ell m}=0$ for $\ell \neq 0$.
In each case, we draw 269 random binary black hole events. 
For the remainder of the manuscript, all analyses focus on the $\lmax=5$ model, as it is not currently computationally practical to include more than the 35 parameters required for $\lmax=5$.
We recover the injected sky distribution, shown in Fig. \ref{fig:injection_recovery}. 
The posterior predictive distribution in the right column closely matches the true distribution, with smaller angular resolution deviations from the true population distribution.
These deviations are not significant.
We report the Bayes factors for each of these scenarios in Table \ref{tab:simulated_event_logbf}.

\bgroup
\def\arraystretch{1.5}
\begin{table}[b!]
    \centering
    \caption{Bayes factors for each of the simulated event analyses.}
    \begin{tabular}{c|c}\hline\hline
        Scenario & $\ln\textrm{BF}^{\textrm{ani}}_{\textrm{iso}}$ \\ \hline
        isotropic & $-33.6$  \\
        dipolar ($a_{11}/a_{00} =0.51$) & $20.0$ \\
        quadrupolar ($a_{22}/a_{00} =0.51$) & $37.6$ \\ \hline\hline
    \end{tabular}
    \label{tab:simulated_event_logbf}
\end{table}
\egroup

In order to study the detectability of anisotropy with a large numbers of events, we simulate thousands of events with perfectly reconstructed sky position, from 100 different realizations of a universe.
By analysing this idealized injection set, we can measure the ``cosmic variance'' in our measurements of the $b_{\ell m}$ ---the uncertainty arising from the finite number of detections.
The total uncertainty, of course, includes contributions from the uncertain sky localization as well as cosmic variance.
Thus, this study yields an optimistic view of what might be possible in the future.

\begin{figure}
    \centering
\includegraphics[width=\linewidth]{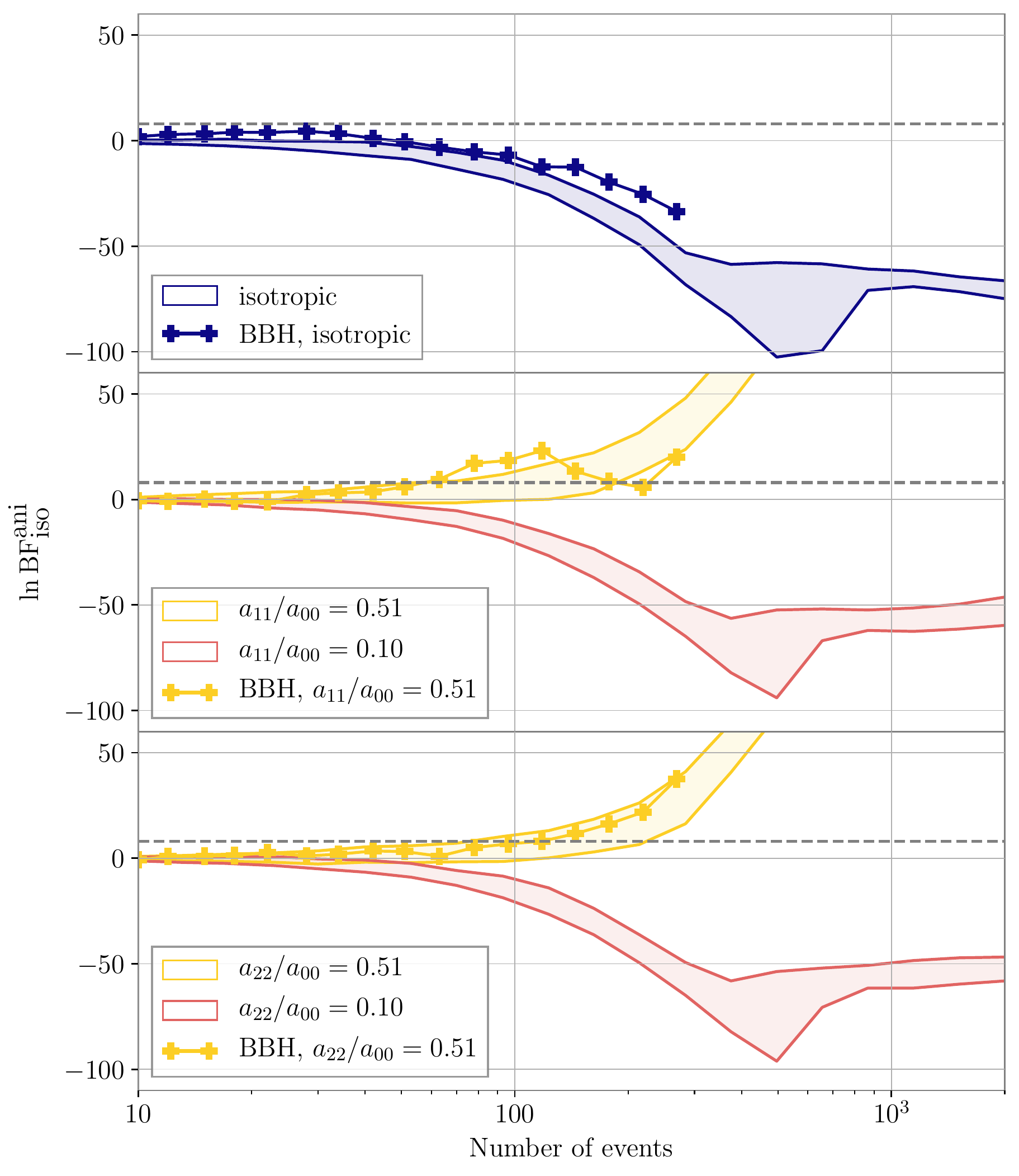}
    \caption{Recovered Bayes factors as a function of the number of binary black hole events observed. 
    The plus markers correspond to the injection study of simulated binary black hole sky positions from full parameter estimation of the source.
    The bounded regions correspond to the $68\%$ confidence intervals for the log Bayes factors recovered from point source estimates. 
    We observe the $\ln\textrm{BF}^\textrm{ani}_\textrm{iso}$ trend for the simulated binary injections to approximately follow the point source estimates. 
    Since isotropy is a compact subset of anisotropic models, the Bayes factor will prefer isotropy until enough events are observed. 
    This is observed in the behavior of the $a_{\ell m}/a_{00} = 0.10$ results.}
    \label{fig:log_bayes}
\end{figure}

\begin{figure}[t!]
    \centering
    \includegraphics[width=\linewidth]{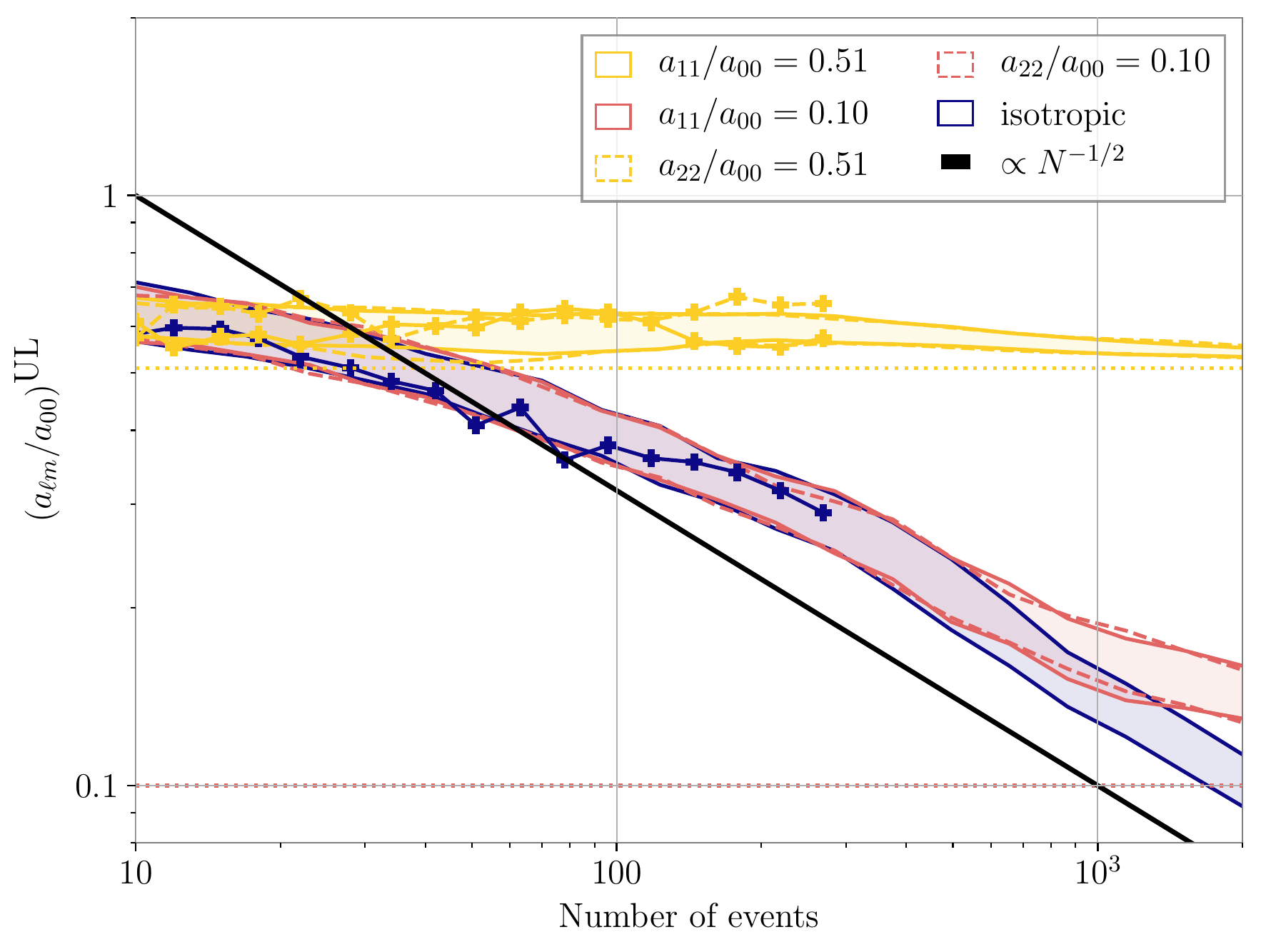}
    \caption{
    Inferred maximum $99.7\%$ upper limits on the anisotropy present in the simulated point source analyses given a number of events.
    The maximum anisotropy is presented for five different scenarios analysed for 100 different realizations with $\lmax=5$ anisotropic models.
    In the regime of a high number of detected events, the maximum anisotropy follows a $\sim N^{-1/2}$ relation. As the number of events increases, the maximum anisotropy measured plateaus towards the true anisotropy.}
    \label{fig:max_aniso}
\end{figure}

In Fig. \ref{fig:log_bayes}, we plot  $\ln\textrm{BF}^\textrm{ani}_\textrm{iso}$ as a function of the number of detections $N$. 
The shaded regions indicate $1\sigma$ confidence intervals.
Since an isotropic model is a subset of the general anisotropic model, as more events are added, the Bayes factor can provide support for the model with a more compact parameter space, until enough evidence mounts for the anisotropic model. 
Therefore, we cannot use a method such as this to ``prove'' isotropy; rather we may provide an increasingly stringent limit on anisotropy. 
In Fig.~\ref{fig:max_aniso} we plot the $99.7\%$ upper limit on $a_{\ell m}/a_{00}$ found in each analysis as a function of the number of events $N$, $(a_{\ell m}/a_{00})^\textrm{UL}$. 
For a low number of events, $(a_{\ell m}/a_{00})^\textrm{UL}$ slowly decreases with the number of events before reaching a consistent power law relation in the high event region.
We find that $(a_{\ell m}/a_{00})^\textrm{UL}$ scales as $\sim N^{-1/2}$ at high number of detected events ($\geq 400$ events). 

\section{Implications}
Looking to the future, as the gravitational-wave network improves in sensitivity, the number of binary black hole events observed will significantly increase. 
Furthermore, the fraction of the those with improved sky localization will also improve \cite{reitze2019cosmic}. 
This will allow for increasing improvements to the estimates of the anisotropy of the binary black hole gravitational-wave sky location distribution. 
Given that upper limit in the estimate of anisotropy scales as $\sim N^{-1/2}$ (Fig. \ref{fig:max_aniso}) in the regime of many events, we can estimate that within one year of operation of third-generation detectors, such as Cosmic Explorer and the Einstein Telescope \cite{reitze2019cosmic, maggiore2019science}, gravitational waves will be able to probe anisotropies at the $\sim0.1\%$ level. 
Coupled with the precise location of sources, this will enable a precise measurement of the anisotropy of the stellar mass binary hole distribution in the universe.
We stress that the inference of (an)isotropy complements (but in no way replaces) the measurements of anisotropy from the cosmic microwave background, which is associated with a very different time in the history of the Universe.
Gravitational-wave observations will provide an additional method to probe the large-scale distribution of stellar mass binary black hole mergers within the Universe.

This analysis provides an approach way to search for the anisotropy of the stellar mass binary black hole sky distribution. 
The results presented in Ref. \cite{stiskalek2020isotropy} qualitatively agree for the predicted population distribution for the sky position of stellar mass binary black holes. 
Our results demonstrate similar probability over-densities with an independent method and parameterization of the anisotropic distribution. 
We also demonstrate the study of both simulated binary black hole mergers and point estimate sources to determine an approximate scaling relation of the inferred maximum anisotropy as a function of the number of detected BBH events.

\section*{Acknowledgements}
We thank Isobel Romero-Shaw for sharing posterior samples for GWTC-1 from the \bilby GWTC-1 catalog paper~\cite{bilby_gwtc1}. 
The authors appreciate the helpful discussions with Colm Talbot and Vuk Mandic. 
We thank Marco Cavaglia and Aditya Vijaykumar for comments on the manuscript.
This work is supported through Australian Research Council (ARC) Centre of Excellence CE170100004. 
EP acknowledges the support of the LSC Fellows program. 
SB is supported by the Doctoral Dissertation Fellowship at UMN. PDL is supported through ARC Future Fellowship FT160100112, and ARC Discovery Project DP180103155.
ET is supported through ARC Future Fellowship FT150100281.
This is document LIGO-P2000212.

This research has made use of data, software and/or web tools obtained from the Gravitational Wave Open Science Center (https://www.gw-openscience.org), a service of LIGO Laboratory, the LIGO Scientific Collaboration and the Virgo Collaboration.
Computing was performed on LIGO Laboratory computing cluster at California Institute of Technology.
We would like to thank all of the essential workers who put their health at risk during the COVID-19 pandemic, without whom we would not have been able to complete this work.

\appendix

\section{GWTC-1 with different models} \label{app:a}
\begin{figure*}
    \centering
    \includegraphics[width=0.9\linewidth]{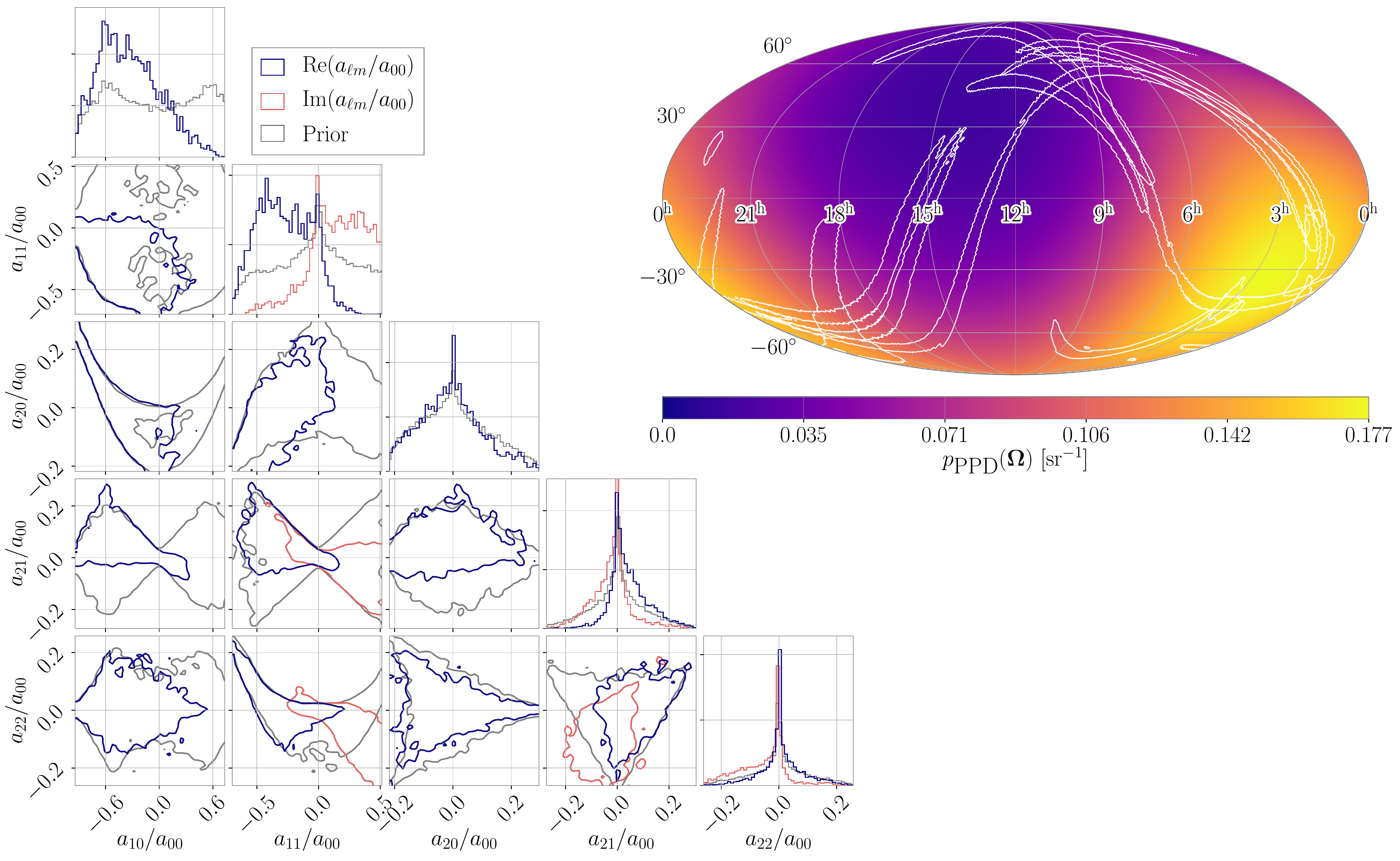}
    \caption{Posterior distribution and posterior predictive distribution for GWTC-1 using a $\lmax=1$ anisotropy model.} \label{fig:lmax1}
\end{figure*}

\begin{figure*}
    \centering
    \includegraphics[width=0.9\linewidth]{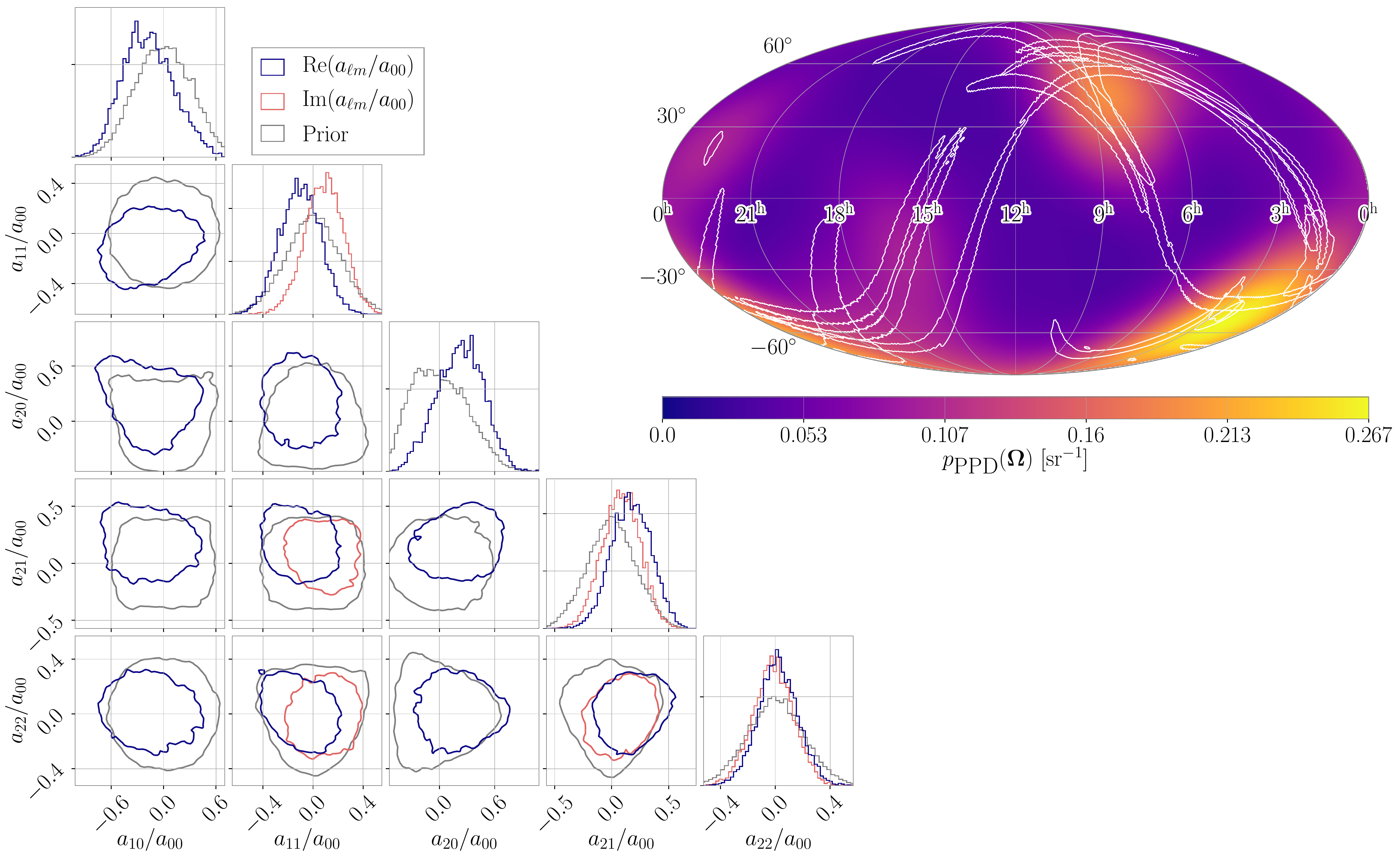}
    \caption{Posterior distribution and posterior predictive distribution for GWTC-1 using a $\lmax=3$ anisotropy model.} \label{fig:lmax3}
\end{figure*}

\begin{figure*}
    \centering
    \includegraphics[width=0.9\linewidth]{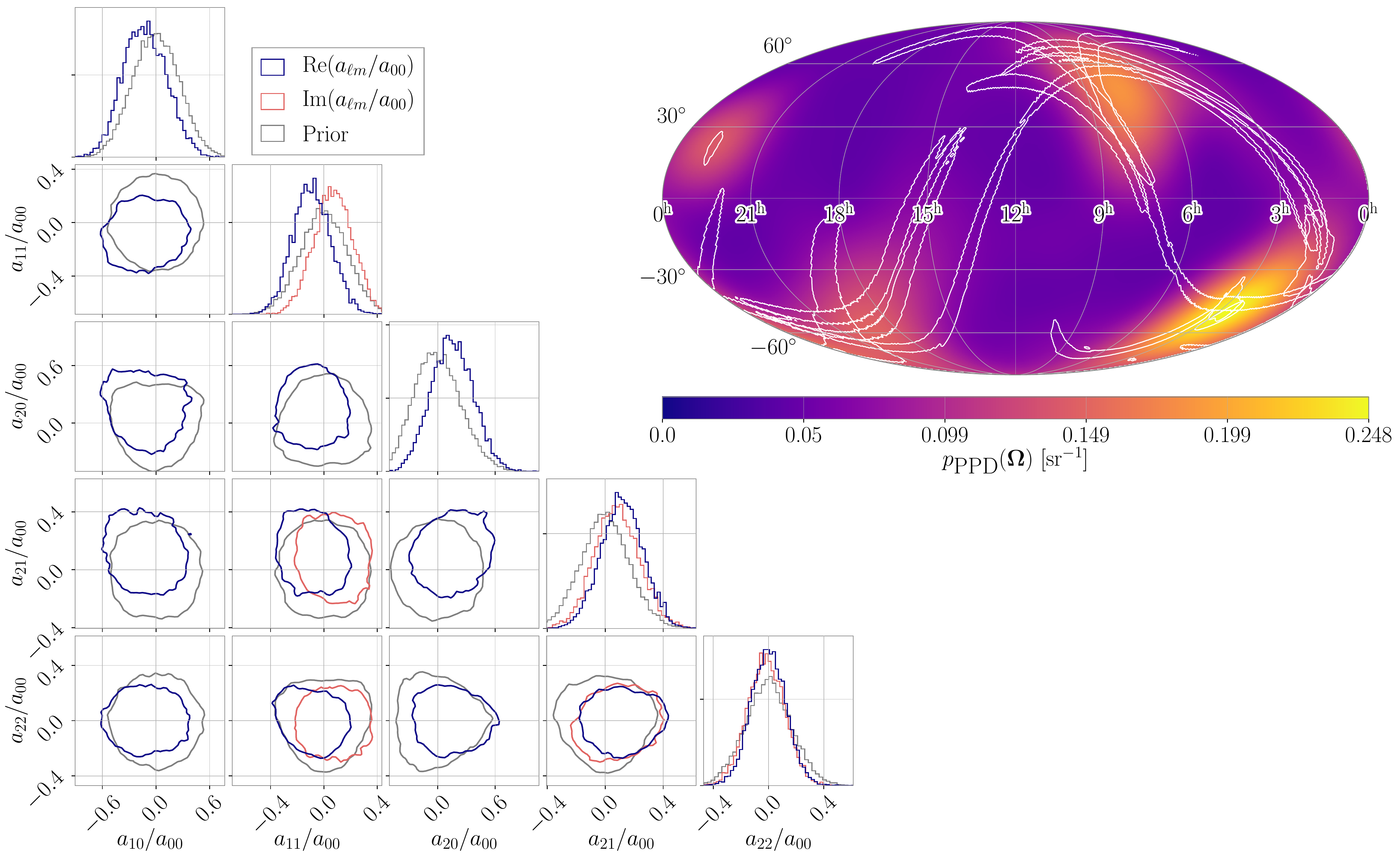}
    \caption{Posterior distribution and posterior predictive distribution for GWTC-1 using a $\lmax=4$ anisotropy model.} \label{fig:lmax4}
\end{figure*}

\begin{figure*}
    \centering
    \includegraphics[width=0.9\linewidth]{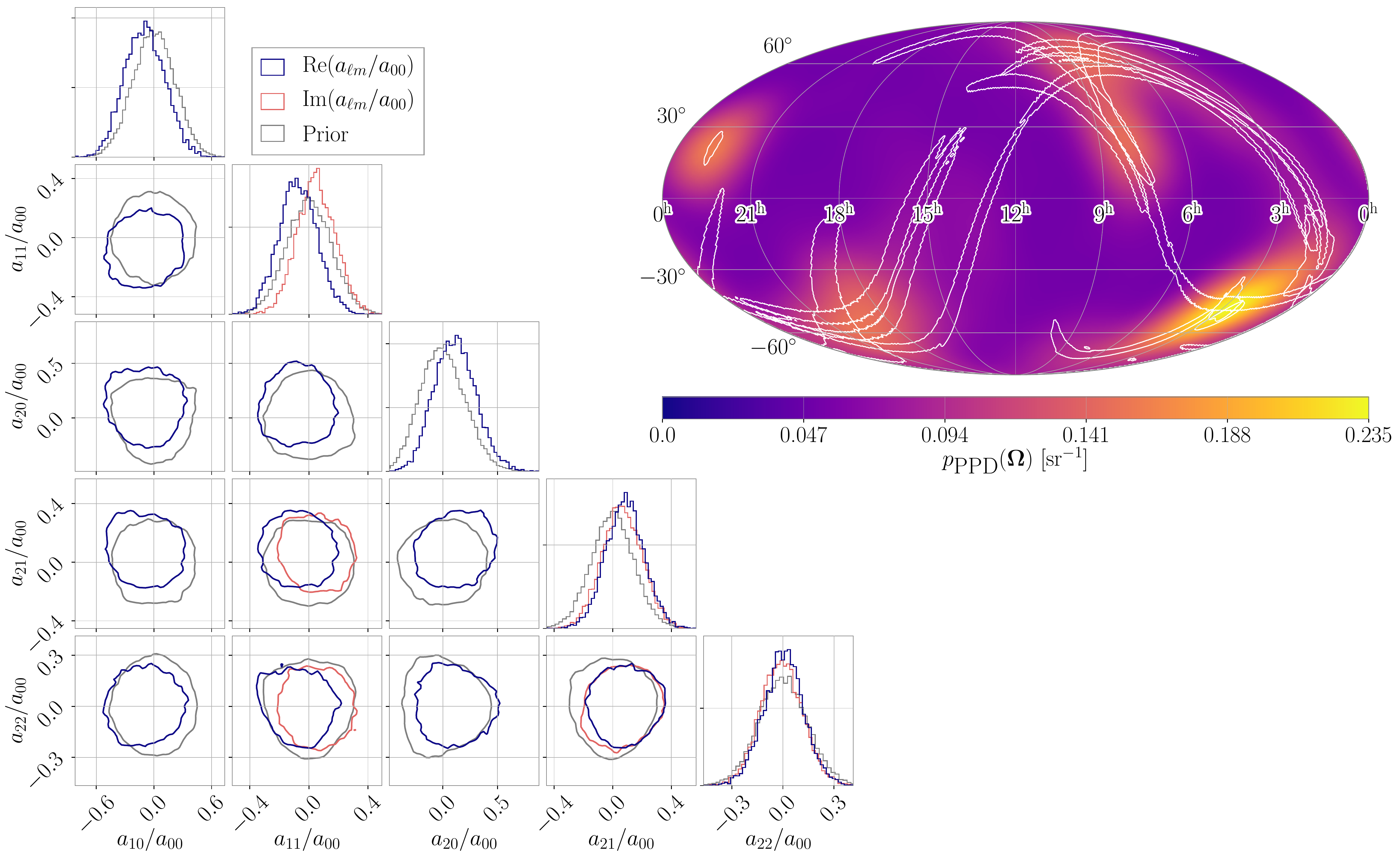}
    \caption{Posterior distribution and posterior predictive distribution for GWTC-1 using a $\lmax=5$ anisotropy model.} \label{fig:lmax5}
\end{figure*}

In Figs. \ref{fig:lmax1}--\ref{fig:lmax5}, we present the additional posterior distributions from the analysis of GWTC-1 undertaken in Sec. \ref{sec:gwtc1}.
Posteriors of $a_{\ell m}/a_{00}$ for $\ell \leq 2$ are shown.
We note that for the $\lmax=1$ model, $a_{2 m}/a_{00}$ posteriors are constructed from Eq. \eqref{eq:alms} which, for any given model, will generate $a_{\ell m}$'s up to $\ell=2\lmax$.
All priors have support at $a_{\ell m}/a_{00} = 0$ for all spherical harmonics up to $\lmax$. 

\bibliography{references}

\end{document}